\documentstyle{elsart}

\begin{document}

\begin{frontmatter}
\title{Explicit solutions of the multi--loop 
integral recurrence relations and its application  \thanksref{paper}}

\thanks[paper]{Talk presented at the AIHENP'96 (EPFL-UNIL 
Lausanne, Sept. 2-6 1996)}

\author[A]{P.\,A.\,BAIKOV \thanksref{AA}}

\thanks[AA]{Supported in part by the RBRF (grant N 96--01--00654),
INTAS (grant 93-1180-ext); e-mail: baikov@theory.npi.msu.su}

\address[A]{ Institute of Nuclear Physics, Moscow State University, 
Moscow~119899, Russia}

\begin{abstract}
The approach to the constructing explicit solutions of the 
recurrence relations for multi--loop integrals are suggested.
The resulting formulas demonstrate a high efficiency,
at least for 3--loop vacuum integrals case.
They also produce a new type of recurrence
relations over the space--time dimension. 
\end{abstract}

\end{frontmatter}

\section{Vacuum case}

Recently \cite{PL,hep-ph2} a new approach to implement recurrence
relations \cite{ch-tk} for the Feynman integrals was proposed.
In this work we extend the general formulas for the solutions
of the recurrence relations to the multi--loop case.
Let us consider first vacuum $L$-loop integrals with $N=L(L+1)/2$ 
denominators (so that one can express through them any scalar product 
of loop momenta) of arbitrary degrees:

\begin{eqnarray}
&&B(\underline{n},D)=
m^{2\Sigma n_i-LD}
\int \cdots \int \frac{d^Dp_1\ldots d^Dp_L} 
{D_1^{n_1}\ldots D_N^{n_N}},
\label{integral}\\
&&D_a=\sum_{i\geq j}A^{(ij)}_a p_i\cdot p_j -\mu_a m^2, \quad
p_k\cdot p_l=\sum_{a=1}^N(A^{-1})_{(kl)}^a(D_a+\mu_a m^2).
\label{den}
\end{eqnarray}

The recurrence relations that result from integration by parts,
by letting  $(\partial/\partial p_i)\cdot p_k$ act on
the integrand \cite{ch-tk}, are:

$$D\delta_k^i B(\underline{n},D)=
2\sum_{a,d=1}^{N}\sum_{l=1}^{L}A_d^{(il)} n_d{\bf I}^{d+}
(A^{-1})_{(kl)}^a({\bf I}^-_a+\mu_a) B(\underline{n},D),
$$

where  
${\bf I}^\pm_c B(\ldots, n_c,\ldots ) = B(\ldots, n_c\pm1,\ldots )$, 
in particular ${\bf I}^\pm_c n_a = n_a\pm \delta^c_a$.
Using the relations

$$[n_d{\bf I}^{d+},{\bf I}^-_a]=\delta_a^d,
\qquad
\sum_{a=1}^{N}A_a^{(il)}(A^{-1})_{(kj)}^a=\delta^{(i}_{(k}\delta^{l)}_{j)},$$

they  can be represented as

\begin{eqnarray}
\frac{D-L-1}{2}\delta_k^i B(\underline{n},D)&=&
\sum_{a,d=1}^{N}\sum_{l=1}^{L}
(A^{-1})_{(kl)}^a({\bf I}^-_a+\mu_a) 
A_d^{(il)} n_d{\bf I}^{d+} 
B(\underline{n},D).
\label{rr1}
\end{eqnarray}

The common way of using these relations is step--by--step reexpression
of the integral (\ref{integral}) with some values of $n_i$ through a set of 
integrals with shifted values of $n_i$, with the final goal to reduce 
this set to a linear combination of several "master" integrals $N_k(D)$
with some "coefficient functions" $F^k(\underline{n},D)$:

$$B(\underline{n},D)=\sum_k F^k(\underline{n},D)N_k(D)$$.

Nevertheless, to find proper combinations of these relations 
and a proper sequence of its use is the matter of art even for
the tree--loop integrals with one mass \cite{REC}. 
Then, even in cases when such procedures were 
constructed, they lead to very time and memory consuming calculation
because of large reproduction rate at every recursion step.
Instead, let us construct the $F^k(\underline{n},D)$ 
directly as solutions of the given recurrence relations. Note, that if
we find any set of the solutions, we could construct $F^k(\underline{n},D)$
as their linear combinations. Let us try the solution of (\ref{rr1}) in the 
following form:

\begin{eqnarray}
f^k(\underline{n})=
\frac{1}{(2\pi\imath)^N}
\oint \cdots  \oint
\frac
{dx_1 \cdots  dx_N}
{x_1^{n_1} \cdots x_N^{n_N}}g(x_a)
\label{solution0}
\end{eqnarray}

where integral symbols denote $N$ subsequent complex
integrations with contours
which will be described later. Acting by some operator
$O_i({\bf I}^-_a, n_a{\bf I}^{a+})$ (all decreasing operators should be 
placed to the left) on
(\ref{solution0}) and performing  the integration by parts one gets
(s.t. are surface terms):

\begin{eqnarray}
O_i({\bf I}^-_a, n_a{\bf I}^{a+})f^k(\underline{n})=
\frac{1}{(2\pi\imath)^N}
\oint \cdot\cdot \oint
\frac
{dx_1 \cdot\cdot dx_N}
{x_1^{n_1} \cdot\cdot x_N^{n_N}}
O_i(x_a, \partial_a)g(x_a)+\mbox{(s. t.)}.\nonumber
\end{eqnarray}

So, if we choose the $g(x_a)$ as the solution of
$O_i(x_a, \partial_a)g(x_a)=0$
and cancel the surface terms by proper choosing of integration contours
(for example, closed or ended in the zero points)
we find that (\ref{solution0}) is a solution of relations 
$O_i({\bf I}^-_a, n_a{\bf I}^{a+})f^k(\underline{n})=0$,
and different choices of contours correspond to different
solutions.

The differential equations for (\ref{rr1}) 
have the solution $g(x_a)=P(x_a+\mu_a)^{(D-L-1)/2}$, where

$$P(x_a)=\det(\sum_{a=1}^N (A^{-1})_{(kl)}^a x_a)$$

is the polynomial in $x_a$ of degree $L$,
so we get the desirable solutions of (\ref{rr1}):

\begin{eqnarray}
f^k(\underline{n},D)=
\frac{1}{(2\pi\imath)^N}
\oint \cdot\cdot \oint
\frac
{dx_1 \cdot\cdot  dx_N}
{x_1^{n_1} \cdot\cdot  x_N^{n_N}}
\det((A^{-1})_{(kl)}^a(x_a+\mu_a))^{\frac{D-L-1}{2}}.
\label{solution}
\end{eqnarray}

Finally, let us derive from (\ref{rr1}) the recurrence relations with 
D-shifts. Note that if $f^k(n_i,D)$ is a solution of (\ref{rr1}), then
by direct substitution to (\ref{rr1}) one can check that 
$P({\bf I}^-_a+\mu_a)f^k(n_i,D-2)$ also is a solution. 
Hence, if $f^k(n_i,D)$ is a complete set of solutions, then

\begin{eqnarray}
f^k(n_i,D)=\sum_n S^k_n(D)P({\bf I}^-+\mu_i)f^n(n_i,D-2),
\nonumber 
\end{eqnarray}

where the coefficients of mixing matrix $S$ is numbers, that is do 
not act on $n_i$. For the solutions (\ref{solution}) the $S$ 
is the unit matrix (the increasing of $D$ by 2 leads to appearing of factor 
$P(x_a)$ in the integrand of (\ref{solution})), but the desire to come to 
some specific set of master integrals may lead to nontrivial mixing.
These relations look different 
from recently proposed in \cite{tarD}, although further investigations
can give some connections with them.

To check the efficiency of this approach we evaluated (using REDUCE)
the first 5 moments in the small $q^2$ expansion of the 3-loop QED 
photon vacuum polarization.
The 3-loop contribution to the moments are expressed through about $10^5$ 
three--loop 
scalar vacuum  integrals with four massive and two massless lines.
The integral (\ref{solution}) in this case can be solved to finite
sums of the Pochhammer's symbols (see \cite{PL}). 
Moreover, it is not necessary to evaluate these integrals separately.
Instead, we evaluated a few integrals of (\ref{solution}) type, but
with $P^{D/2-2}$ producted by a long polynomial in $x_i$
(the results see in \cite{PL,3l}, they are in agreement
with QCD calculations \cite{CKS} made by FORM).

The comparison with the recursive approach shows a reasonable 
progress: the common way used in \cite{3l} demands  
several CPU hours on DEC-Alpha to calculate full $D$ 
dependence of the first moment, and further calculations became 
possible only after truncation in $(D/2-2)$. In the present approach the 
full $D$ calculation for the first moment demands a few minutes on PC.

\section{Non--vacuum case}

Suppose that
integrals (\ref{integral}) depend on $R$ external momenta $p_i$
($L < i \leq L+R$). The number of the denomenators are now 
$N_1=L(L+1)/2+LR$,
and the number of additional ("external") invariants are $N_2=R(R+1)/2$.
Let us expand the integrals in formal seria over 
"denominator--like" objects $D_a$ of (\ref{den}) type with  
$a=N_1+1,..,N_1+N_2$, depending on external momenta only:

$$B(n_{l, (l=1,\dots, N_1)}, p_{k, (k=L+1,\dots,L+R)})=
\int \cdots \int \frac{d^Dp_1\ldots d^Dp_L}{D_1^{n_1}\ldots
D_{N_1}^{n_{N_1}}}=$$
\begin{equation}
=\sum_{{n_i}(i>N_1)}m^{-2\Sigma n_i+2N_2+LD}b(n_i,_{(i=1,\dots, N_1+N_2)})
\prod_{i=N_1+1}^{N_1+N_2}
D_i^{n_i-1}.
\nonumber 
\label{integral1}
\end{equation}

We define such general expansion in order to write 
the recurrence relations in compact form, in practice
the coefficients $A^{(ij)}_a$ and $\mu_a$ may be very simple. 
The expansion with negative $n_i$ corresponds to the large momenta 
expansion, with positive ones to the expansion near points $\mu_a\,m^2$. 
The $n_i$ can also be noninteger, but with unit shifts.

Acting by
$(\partial/\partial p_i)\cdot p_k$,  $(i=1,\dots,L; k=1,\dots,L+R)$
on the integrand we get $N_1$ recurrence relations.
The additional $N_2$ relations we get acting by
$p_k \cdot (\partial/\partial p_i)$, $(i,k=L+1,\dots,L+R)$ on
both sides of (\ref{integral1}).
These new relations look like the old ones with only exception that they
have no terms proportional to space--time dimention $D$.
The complete set of recurrence relations is now

$$((D-L-R-1)\,\delta_k^i-(D-R-1)\,\hat{\delta_k^i})
\,b(\underline{n},D)= \qquad
$$
$$
\qquad
=2\sum_{a,d=1}^{N_1+N_2}\sum_{l=1}^{L+R}
(A^{-1})_{(kl)}^a({\bf I}^-_a+\mu_a) 
A_d^{(il)} n_d{\bf I}^{d+}
b(\underline{n},D),
$$

where $\hat{\delta_k^i}$=($\delta_k^i$ if $i,k> L$, else 0).
The corresponding differential equations have the solution
$g(x_a)=g'(x_a+\mu_a)$, where

\begin{equation}
g'(\underline{x})=\det\Bigl((A^{-1})_{(kl)}^a 
x_a\Bigr)^{\frac{D-L-R-1}{2}}
{\det}_0\Bigl((A^{-1})_{(kl)}^a 
x_a\Bigr)^{-\frac{D-R-1}{2}},
\label{solution1}
\end{equation}

and $\det_0$ denotes the minor with $k,l>L$.
So, one can use the representation (\ref{solution0}),
but the problem of resolving it to explicit formulas demands futher 
investigations.

Finally note that one can formally obtain the formulas (\ref{solution},
\ref{solution1}) by "change of integration variables" from
loop momenta to "denomenator--like objects" $D_a$. The 
weight function for this change is 

$$
\int d^Dp_1\cdot\cdot d^Dp_L 
\prod_i \delta(D_i/m^2-x_i)
\propto \det((A^{-1})_{(kl)}^a(x_a+\mu_a))^{\frac{D-L-1}{2}}.
$$

\end{document}